\documentclass[twocolumn]{article}

\usepackage[utf8]{inputenc}
\usepackage[margin=2.5cm]{geometry}
\usepackage{authblk}
\usepackage{setspace}
\usepackage{graphicx}
\graphicspath{{./figures/}}
\usepackage{subcaption}
\usepackage{amsmath}
\usepackage{booktabs}
\usepackage{hyperref}
\usepackage{placeins}
\usepackage{float}
\usepackage[numbers]{natbib}

\usepackage[switch]{lineno}       % 'switch' puts line numbers on outer margins
\title{Development of a Simple Stellarator using Tilted Circular Toroidal Field Coils }

\author[1*$\dag$]{Ashit Kumar Nath}
\author[1$\dag$]{Yasuhiro Suzuki}
\affil[1]{Graduate School of Advanced Science and Engineering, Hiroshima University, Higashi-Hiroshima 739-8527, Japan}
\date{}

\begin{document}
\twocolumn[
\begin{@twocolumnfalse}

\maketitle

%%%%%% Abstract (two-column) %%%%%%
\begin{abstract}
This study investigates a simplified stellarator configuration employing circular coils, in which rotational transform is generated by tilting the toroidal field (TF) coils. A pair of axisymmetric poloidal field (PF) coils is introduced to compensate for the vertical magnetic field component produced by the tilted TF coils, together forming the three-dimensional magnetic configuration. The existence of clear, nested magnetic flux surfaces is confirmed through magnetic field-line tracing, and the corresponding vacuum free-boundary equilibrium is computed using the DESC solver. The coil set is partially optimized by varying the TF coil radius and tilt angle to reduce neoclassical transport and enhance alpha-particle confinement. The optimized configuration is compared with fully optimized stellarators such as W7-X and LHD in terms of alpha-particle confinement and the $\Gamma_C$ proxy. The neoclassical transport coefficient $D_{11}$ is evaluated and found to be low. Collisionless guiding-center orbit calculations for 100~eV protons and 3.5~MeV alpha particles further demonstrate favorable confinement properties.
\vspace{0.5cm}
\newline
Keywords: Simple stellarator, Circular tilted coil, Effective ripple, Alpha particle confinement, GammaC 
\end{abstract}

\vspace{1em}
\end{@twocolumnfalse}
]

%%%%%% Main Text %%%%%%

\section{Introduction}
The stellarator is a magnetic confinement fusion device that uses complex, three-dimensional magnetic fields generated entirely by external coils to confine high-temperature plasma for sustained fusion reactions, whereas the tokamak is an axisymmetric device in which confinement is achieved through externally applied toroidal magnetic fields and a strong induced toroidal plasma current. Although tokamaks demonstrate excellent confinement performance, their reliance on a large plasma current can trigger current-driven instabilities and severe disruptions in high-confinement regimes such as H-mode. Additionally, MHD instabilities, including edge localized modes (ELMs)can generate transient heat loads on plasma-facing components, posing challenges for reactor operation~\cite{Wesson78,Zohm96}.
 Unlike tokamaks, stellarators do not rely on a toroidal plasma current for confinement, enabling steady state operation and reducing the risk of plasma disruptions, thereby offering improved operational stability for long-duration fusion applications~\cite{Boozer98,Helander14}. However, transport properties, particularly neoclassical transport of stellarators, are poor due to large helical ripple components in the magnetic field~\cite{HoKulsrud87}. To mitigate these issues, optimized stellarator configurations have been developed, and their improved confinement properties have been validated both experimentally and theoretically in devices such as the Large Helical Device (LHD), Wendelstein 7-X (W7-X), and the Helically Symmetric Experiment (HSX)~\cite{Boozer98, Grieger89,Anderson95,Klinger17}. These devices rely on sophisticated modular and helical coils to generate optimized magnetic fields, leading to high costs and engineering complexity, which have historically limited the scalability of stellarator reactors compared to tokamaks. A notable example is the National Compact Stellarator Experiment (NCSX) in the United States, which was canceled in 2008 due to significant delays and cost overruns associated with the extreme precision required for manufacturing and assembling its complex modular coils~\cite{Chrzanowski09}. Therefore, the feasibility of extending these sophisticated coil designs to future reactor-scale devices remains a critical issue, motivating efforts toward developing simpler coil geometries in stellarator design.
Throughout the history of stellarator research, configurations based on simple coil sets have been investigated by many researchers~\cite{Bykov89, Moroz95, PedersenBoozer02, Clark14, Suzuki21}. Early studies~\cite{Bykov89, Moroz95} confirmed that tilted toroidal field coils can generate rotational transform and form nested magnetic flux surfaces. The Columbia Nonneutral Torus (CNT)~\cite{PedersenBoozer02} employs four circular coils to create a two-period, low–aspect-ratio stellarator and demonstrated well-defined flux surfaces with good electron confinement, showing that effective stellarator fields can be produced without complex non-planar coils. Similarly, Proto-CIRCUS~\cite{Clark14}, a small tokamak–torsatron hybrid device, uses six tilted circular coils to generate rotational transform, illustrating that planar coil geometries can achieve stellarator-like confinement with reduced plasma current and magnetic ripple. More recently, Suzuki \textit{et al.}~\cite{Suzuki21} proposed a simple stellarator configuration employing sixteen rectangular tilted toroidal field coils, and demonstrated the formation of clear nested flux surfaces with low helical ripple and quasi-isodynamic-like magnetic properties. However, detailed studies of neoclassical transport and fast-ion confinement, especially alpha-particle confinement, remain largely absent in simple-coil stellarators.

This study, therefore, investigates the partial optimization of a tilted circular coil stellarator, following ~\cite{Moroz95,Suzuki21}, aiming to obtain closed flux surfaces of significant volume, proper confinement within the coil system, and adequate vacuum rotational transform, while simultaneously achieving low neoclassical transport and enhanced fast-ion confinement. The paper is structured as follows. The generation of nested flux surfaces and the optimization procedure are described in section~\ref{sec:methodology}. The neoclassical transport and fast-ion confinement results are presented in section~\ref{sec:results}. Finally, section~\ref{sec:conclusion} provides a summary of the study.

\section{Methodology}
\label{sec:methodology}
\subsection{Realization of the Magnetic Field}
\begin{figure}[tb] \centering \includegraphics[width=7cm,clip]{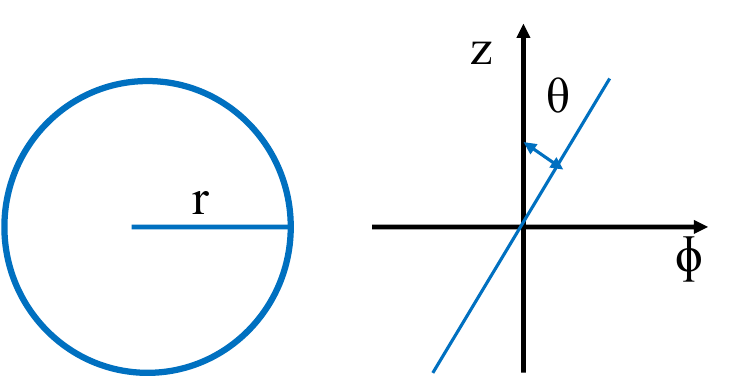} \caption{Schematic illustration of a tilted toroidal field (TF) coil showing (a) the $R$–$Z$ plane and (b) the $\phi$–$Z$ plane.} \label{fig:tfcoil_schematic} \end{figure}

\begin{figure}[tb]
  \centering
  \includegraphics[width=7cm,clip]{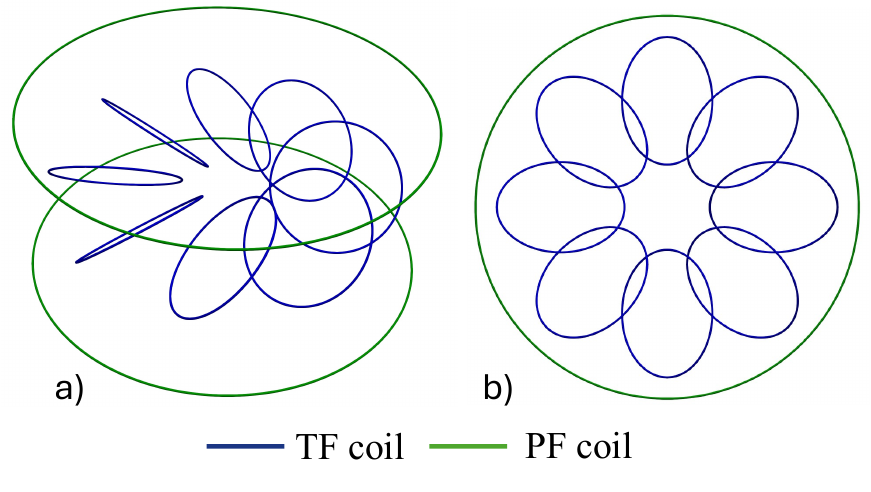}
  \caption{(a)Isometric and (b)top view of the stellarator coil set geometry. }
  \label{fig:coilset_geometry}
\end{figure}

The vacuum magnetic field was generated using tilted circular TF coils. Fig.~\ref{fig:tfcoil_schematic} shows a coil of radius $r$ tilted by $\theta$, centered at $R_0 = 1~\mathrm{m}$. The full configuration shown in Fig.~\ref{fig:coilset_geometry} comprises eight TF coils and a pair of PF coils. The TF coil current is estimated using~[1].

\begin{equation}
I = \int_S \mathbf{J} \cdot \hat{n} \, d^2x 
  = \frac{1}{\mu_0} \oint_{\partial S} \mathbf{B} \cdot d\mathbf{l} 
  = \frac{2\pi R B_\phi}{\mu_0}.
\label{eq:tf_current}
\end{equation}

\begin{figure}[h!]
\centering
\includegraphics[width=5cm,clip]{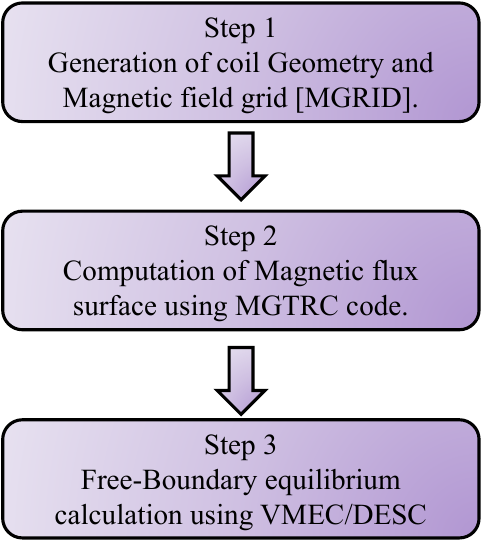}
\caption{Methodology flow for determining free-boundary stellarator equilibria.}
\label{fig:method-flow}
\end{figure}

\begin{figure*}[t]
  \centering
  \includegraphics[width=0.9\textwidth,clip]{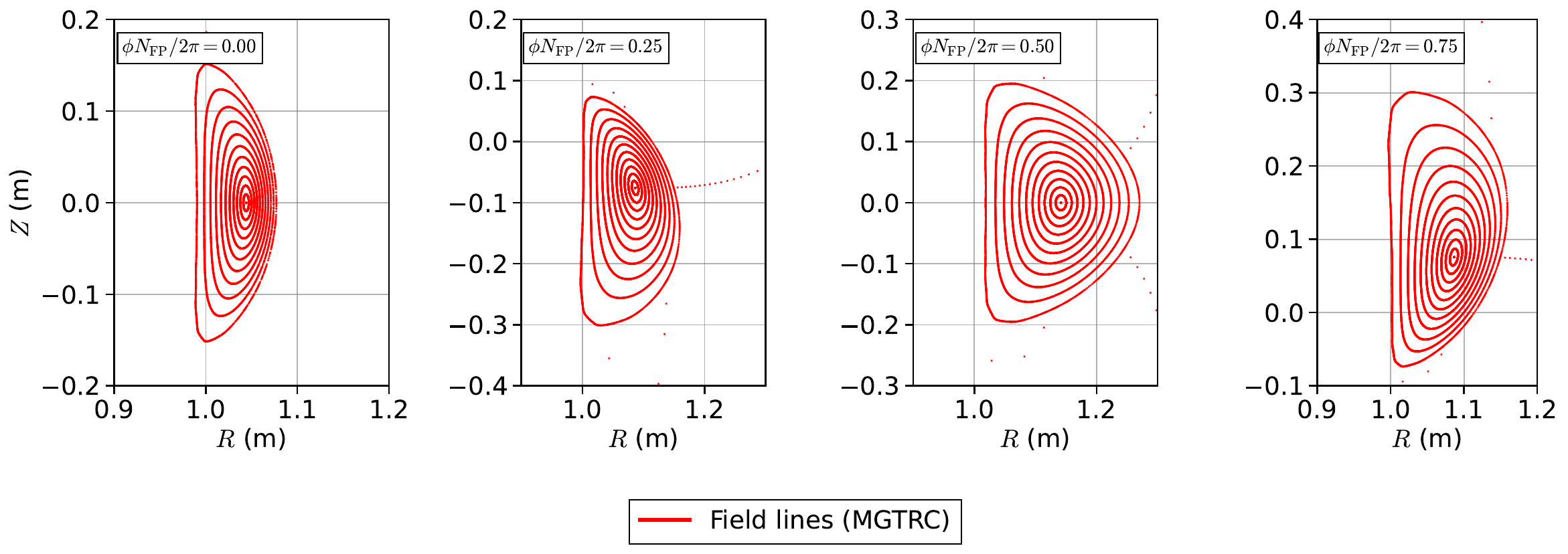}
  \caption{Poincaré plots obtained from magnetic field-line tracing using theMGTRC code, demonstrating the presence of well-defined nested magnetic flux surfaces. The cross-sections are shown at four representative toroidal locations, $\phi N_{\mathrm{FP}}/2\pi = 0.0$ ,$0.25$, $0.5$ and $0.75$, in cylindrical $(R,Z)$ coordinates.}
  \label{fig:mgtrc}
\end{figure*}

Assuming the normalization condition  $R_0 B_\phi = 1$ [Major radius $R_0 = 1$  meter and Magnetic field $B_\phi = 1$ Tesla], the toroidal field coil (TF) currents for configurations with NFP $= 8$ [no of field periods] are calculated to be  $625~\mathrm{kA}$.
The overall numerical workflow employed to obtain free-boundary stellarator equilibria is summarized in Fig.~\ref{fig:method-flow}. First, the coil geometry was specified in a text-based coil file, which the MAKEGRID code used to generate the MGRID file. The MAKEGRID code uses the Biot–Savart law to compute the magnetic-field contribution of each coil segment on a three-dimensional grid, thereby defining the vacuum magnetic configuration. Second, the presence of nested magnetic flux surfaces was verified using the MGTRC (magnetic field-line tracing) code ~\cite{MGTRC}, as shown in Fig. \ref{fig:mgtrc}. In addition, the PF-coil current and the toroidal magnetic flux $\psi_T$
(PHIEDGE in VMEC; \texttt{eq.psi} in DESC) were accurately estimated using
the MGTRC code, which then served as input for the free-boundary
equilibrium calculation.

\begin{figure*}[t]
  \centering
  \includegraphics[width=0.9\textwidth,clip]{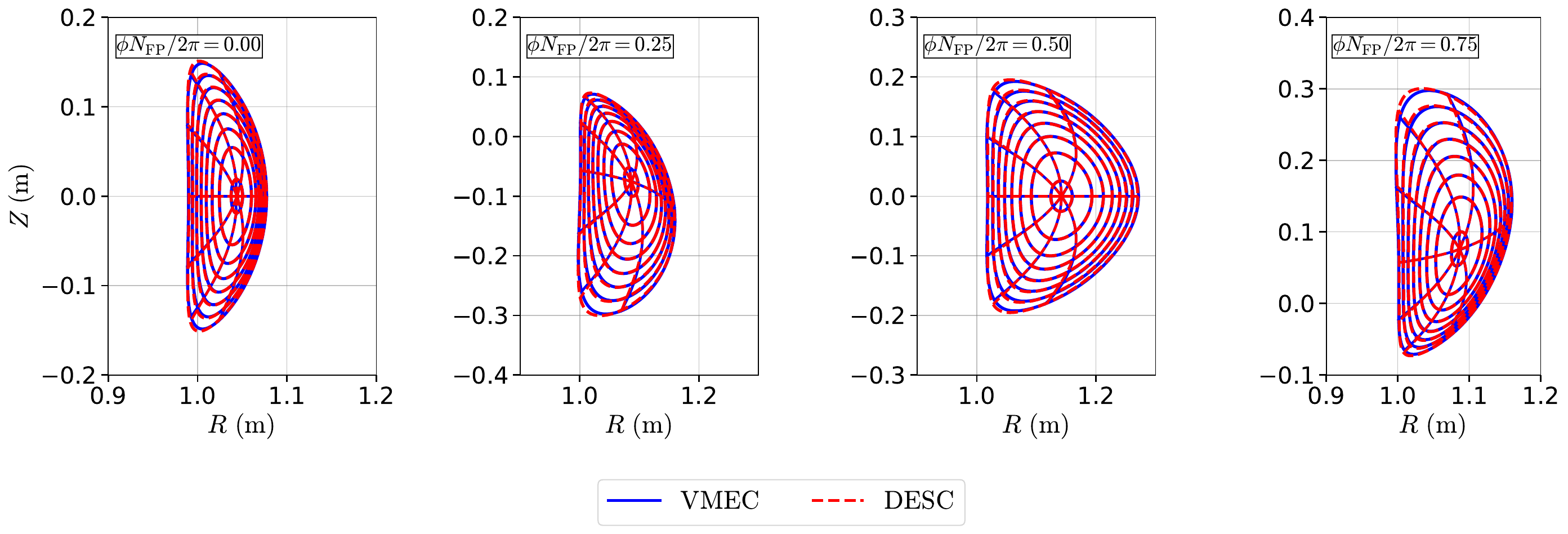}
  \caption{Comparison of magnetic flux surfaces computed using \textsc{VMEC} (solid blue) and \textsc{DESC} (dashed red) at the same toroidal locations. The close agreement between the two equilibrium solvers confirms consistent reconstruction of nested magnetic flux surfaces, in agreement with the MGTRC field-line tracing results shown in figure \ref{fig:mgtrc}}
  \label{fig:vmec_desc}
\end{figure*}

\subsection{Calculation of Vacuum Free-Boundary Equilibrium}

Once the presence of nested magnetic flux surfaces along with the corresponding toroidal field (TF) and poloidal field (PF) coil currents was confirmed, the vacuum free-boundary equilibrium was computed using the \textsc{DESC}[8]. In this work, \textsc{DESC} was used as the primary equilibrium solver; however, \textsc{VMEC} was also employed where necessary to verify the
accuracy and quality of the generated magnetic flux surfaces. A direct comparison between the flux surfaces obtained from
\textsc{DESC} and \textsc{VMEC} is shown in Fig.~\ref{fig:vmec_desc}, demonstrating excellent
agreement in the nested magnetic flux surface geometry, consistent with the flux surfaces generated using the MGTRC field-line tracing code
shown in Fig.~\ref{fig:mgtrc}. Both \textsc{VMEC} and \textsc{DESC} compute three-dimensional ideal
magnetohydrodynamic (MHD) equilibria. Both codes seek to satisfy the
fundamental MHD equilibrium conditions given in Eq.~(2), but differ in
their mathematical formulation and numerical implementation.
\begin{equation}
\mathbf{J} \times \mathbf{B} = \nabla p,
\qquad
\nabla \times \mathbf{B} = \mu_0 \mathbf{J},
\qquad
\nabla \cdot \mathbf{B} = 0.
\end{equation}

\begin{equation}
W = \int_V \left( \frac{|\mathbf{B}|^2}{2\mu_0}
+ \frac{p}{\gamma - 1} \right)\, \mathrm{d}V,
\end{equation}
\textsc{VMEC} computes three-dimensional equilibria by minimizing the
total plasma energy function, subject to the constraint of nested magnetic flux surfaces \cite{HirshmanWhitson83}.
The equilibrium is obtained using a steepest-descent algorithm, with the
radial dependence discretized via finite differences and the angular
dependence represented using a Fourier series. While this variational approach is robust and widely used, the finite-difference treatment in the radial direction can limit accuracy, particularly in the vicinity
of the magnetic axis.
\begin{equation}
\mathbf{F}(\mathbf{r}) = \mathbf{J} \times \mathbf{B} - \nabla p,
\end{equation}
In contrast, \textsc{DESC} adopts a force-based formulation and directly
minimizes the local ideal MHD force residual,
evaluated at collocation points throughout the plasma volume \cite{DudtKolemen20}. The equilibrium equations in DESC are solved using a Newton–Raphson type method combined with a global Fourier-Zernike pseudospectral representation of
the spatial coordinates. This formulation provides analytic spatial derivatives and exponential convergence in the radial direction, enabling
high-accuracy solutions even near the magnetic axis. As demonstrated by Panici \text{et al} \cite{Panici23}, this approach allows \textsc{DESC} to achieve a specified force-balance tolerance with significantly reduced computational cost compared to \textsc{VMEC}, while maintaining mutual consistency between the two solvers for
configurations with well-nested magnetic flux surfaces. Therefore, DESC is used throughout this study for the equilibrium calculations.

\subsection{Optimization Procedure}

\begin{table}[h!]
\centering
\caption{Parameter scan space for TF coil optimization.}
\begin{tabular}{l p{4cm}}
\hline
\textbf{Parameter} & \textbf{Values} \\
\hline
Tilt angle [deg]   & 30, 35, 40, 45, 50 \\
TF coil radius [m] & 0.25, 0.30, 0.35, 0.40, 0.45, 0.50, 0.55, 0.60, 0.65 \\
\hline
\end{tabular}
\label{tab:coil_scan}
\end{table}

The optimization of the coil system was carried out by computing a series of vacuum-free-boundary equilibria for a wide range of toroidal-field (TF) coil geometries, while keeping the poloidal-field (PF) coil set fixed. Throughout this study, the PF coils maintain a constant geometry with a radius of $1.8\,\mathrm{m}$ and vertical positions at $Z=\pm 0.8\,\mathrm{m}$. Two geometric parameters of the TF coils are systematically varied. The TF coil tilt angle was scanned from $30^\circ$ to $50^\circ$ in increments of $5^\circ$, while the coil radius was varied from $0.25\,\mathrm{m}$ to $0.65\,\mathrm{m}$ in steps of $0.05\,\mathrm{m}$. This parameter space results in a total of 45 distinct TF coil configurations (5 tilt angles $\times$ 9 radii) for the $\mathrm{NFP}=8$ stellarator configuration, as summarized in Table~\ref{tab:coil_scan}. The entire parameter scan is automated through a Python-based workflow that interfaces with \texttt{DESC} to compute vacuum-free-boundary equilibria for each coil geometry. For every configuration, the corresponding toroidal-field (TF) and poloidal-field (PF) coil currents, together with the toroidal magnetic flux $\psi_T$ (accessed as \texttt{eq.psi} in \texttt{DESC}), are first determined following the procedure described in Sec.~2.1. Once the equilibria were obtained, key parameters---including the effective helical ripple (or effective ripple) $\varepsilon_{\mathrm{eff}}$, neoclassical transport coefficients such as $D_{11}$, and alpha-particle confinement---were evaluated. The results of this systematic analysis are presented in Sec.~\ref{sec:results}.

\begin{figure*}[h!]
  \centering
  \includegraphics[width=0.7\textwidth,clip]{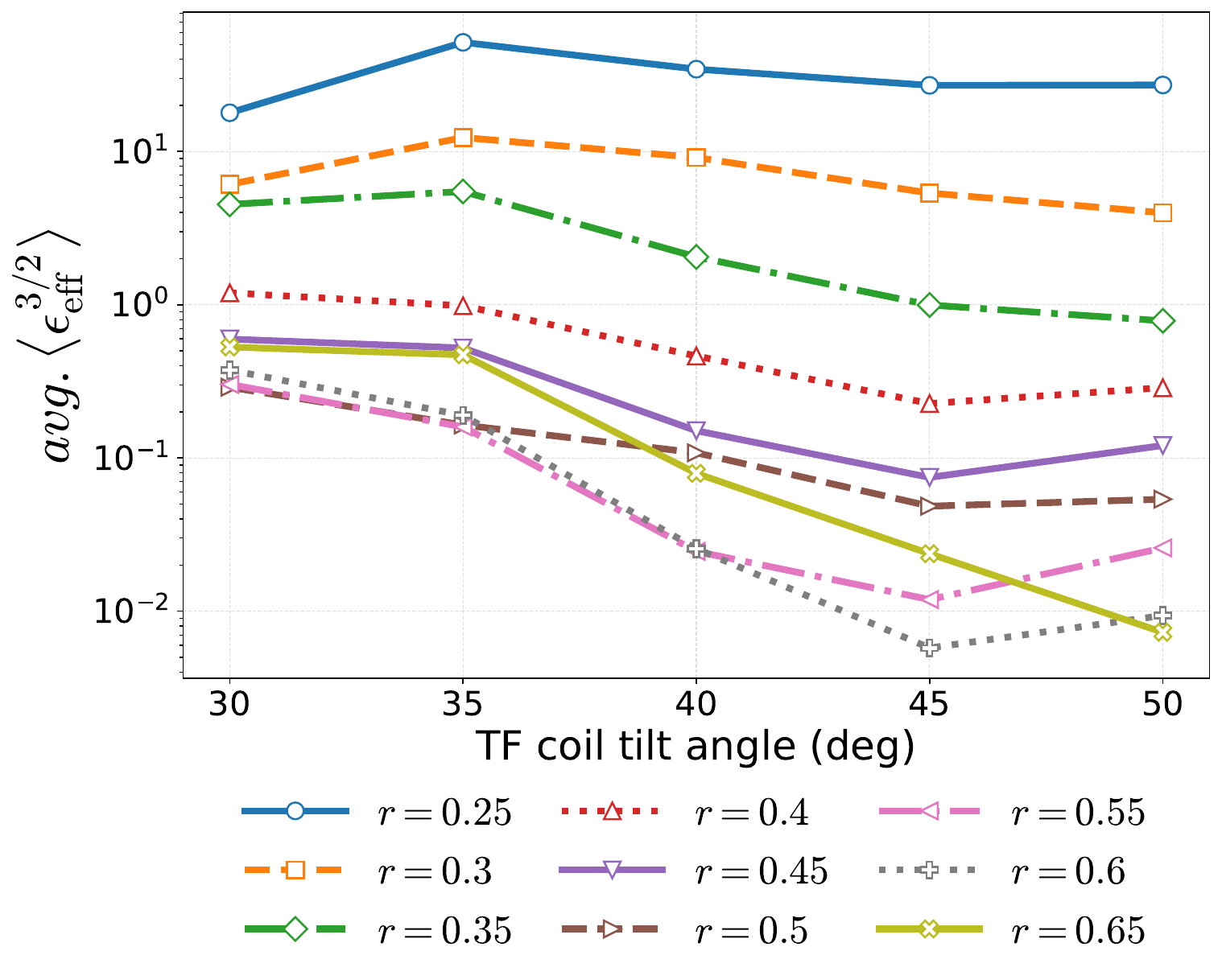}
  \caption{Variation of averaged effective ripple
  $\langle \epsilon_{\mathrm{eff}}^{3/2} \rangle$ with TF coil tilt angle for the
  8-NFP stellarator configuration under vacuum magnetic field conditions
  ($\beta = 0$). Each curve corresponds to a distinct TF coil radius $r$. [refer to Appendix Fig.~\ref{fig:6_eff} and Fig.~\ref{fig:10_eff} for 6-NFP and 10-NFP]}
  \label{fig:eff} 
\end{figure*}

\section{Results and Discussion}
\label{sec:results}
\subsection{Effective Ripple}
\label{subsec:effective-ripple}
The neoclassical effective ripple, $\epsilon_{\mathrm{eff}}^{3/2}$, serves as a quantitative measure of magnetic field non-uniformity in stellarators and is directly linked to neoclassical transport in the $1/\nu$ regime~\cite{Nemov99}. Lower values of $\epsilon_{\mathrm{eff}}$ correspond to a smoother magnetic field strength along particle orbits and therefore to reduced neoclassical transport losses and improved particle confinement. In this work, the effective ripple was computed using the NEO code ~\cite{Nemov99}, which evaluates monoenergetic transport coefficients based on the full three-dimensional magnetic field structure. The required magnetic field information was provided in the form of \text{boozmn.nc} files, generated through a Boozer coordinate transformation~\cite{Boozer81} of the free-boundary equilibrium obtained from the equilibrium solver (DESC/VMEC). All calculations were performed assuming a vacuum magnetic field ($\beta = 0$).
Fig.~\ref{fig:eff} shows the variation of the averaged effective ripple, $\langle \epsilon_{\mathrm{eff}}^{3/2} \rangle$, as a function of the toroidal field (TF) coil tilt angle. The TF coil tilt angle is shown on the horizontal axis, while the vertical axis represents the averaged effective ripple, $\langle \epsilon_{\mathrm{eff}}^{3/2} \rangle$,
plotted on a logarithmic scale. Each curve corresponds to a different TF coil
radius, as indicated in the legend.
The averaged effective ripple is computed as the arithmetic mean of the radial profile obtained from the NEO code,
\begin{equation}
\langle \epsilon_{\mathrm{eff}}^{3/2} \rangle
=
\frac{1}{N}
\sum_{i=1}^{N}
\epsilon_{\mathrm{eff}}^{3/2}(\rho_i),
\end{equation}
where $\epsilon_{\mathrm{eff}}^{3/2}(\rho_i)$ denotes the effective ripple
evaluated at the $i$-th normalized radial position $\rho_i$, and $N$ is the
total number of radial grid points.

\begin{figure*}[h!]
  \centering
  \includegraphics[width=0.7\textwidth,clip]{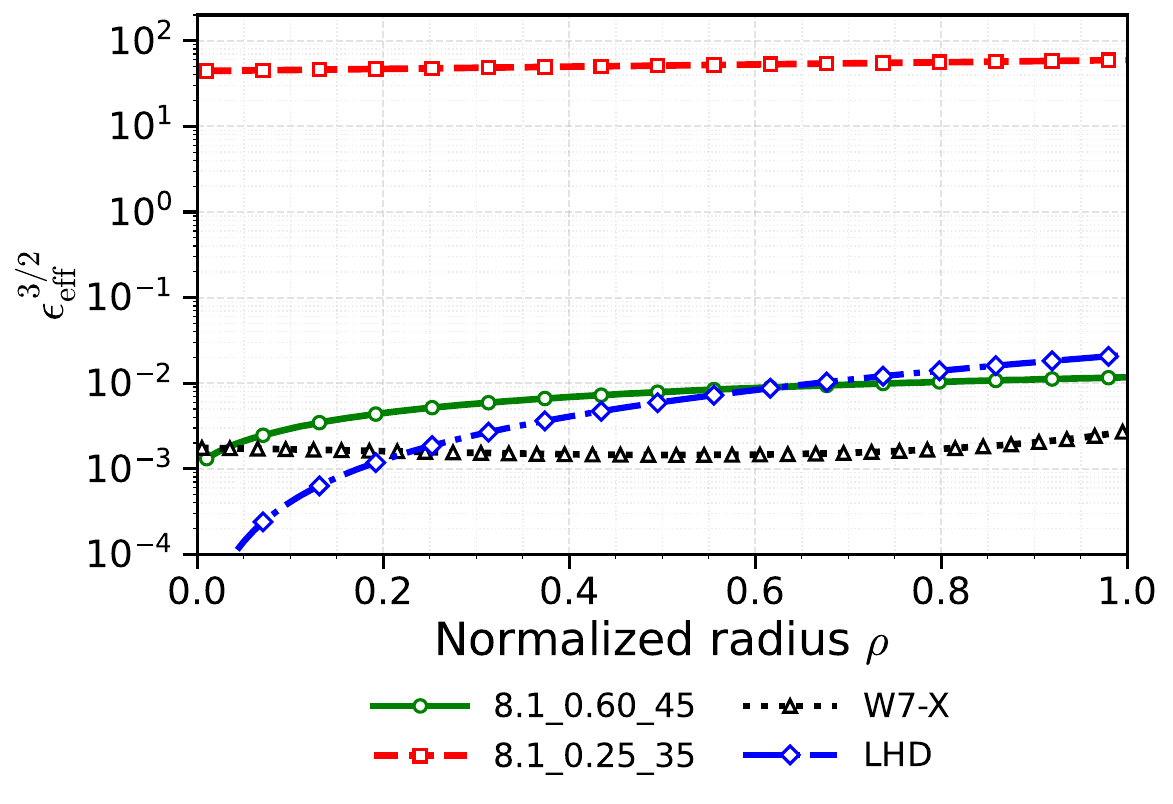}
  \caption{Radial profiles of the neoclassical effective ripple $\epsilon_{\mathrm{eff}}^{3/2}$ plotted versus the normalized radius $\rho$ for two 8-TF simple-coil configurations, compared with the W7-X and LHD reference cases.}
  \label{fig:eff_ripple_comp}
\end{figure*}

As shown in Fig.~\ref{fig:eff}, increasing the TF-coil tilt angle leads to an
overall reduction of the effective ripple. In addition to its dependence on the TF-coil tilt angle, the effective ripple also shows a strong dependence on the TF-coil radius. For a given tilt angle, the averaged effective ripple generally decreases with increasing TF-coil radius up to
$r \approx 0.6$, beyond which it increases again at $r = 0.65$. This behavior
indicates that the minimum effective ripple is achieved near $r \approx 0.6$,
suggesting an optimal radial location where magnetic field non-uniformity is
minimized. The increase in effective ripple at $r = 0.65$ indicates a degradation in magnetic field optimization beyond the optimal coil radius, suggesting that excessively large TF-coil radii introduce unfavorable field non-uniformities possibly arising from increased magnetic field variations near the plasma boundary.

An optimal TF-coil tilt range is observed around $40^\circ$--$45^\circ$, where
effective ripple reaches its minimum for most
radial locations. Beyond this range, further increases in the tilt angle lead to
saturation or a slight degradation of the effective ripple for certain flux
surfaces, indicating no significant improvement. Notably, the configuration with radius of $r \approx 0.6$ and a
TF-coil tilt angle of $45^\circ$ exhibits an effective ripple of
$\langle \epsilon_{\mathrm{eff}}^{3/2} \rangle < 10^{-2}$. Such a low level of effective ripple is generally considered favorable for stellarator operation, as it corresponds to significantly reduced neoclassical transport losses and
improved particle confinement.

To further elucidate the dependence of the neoclassical effective ripple on the
TF-coil geometry, Fig.~\ref{fig:eff_ripple_comp} compares the radial profiles of
$\epsilon_{\mathrm{eff}}^{3/2}$ for stellarator configurations exhibiting the
lowest and highest effective ripple, as identified from Fig.~\ref{fig:eff}.
The specific configurations selected for this comparison are summarized in
Table~\ref{tab:config_comparison}. Each configuration is labeled using the convention
(\textit{number of TF coils}).(\textit{number of PF coil pairs})\_(\textit{TF coil radius in meters})\_(\textit{TF coil tilt angle in degrees}).

\begin{table}[h!]
\centering
\caption{Configurations with high and low effective ripple.}
\label{tab:config_comparison}

\begin{tabular}{c c c}
\toprule
\textbf{NFP} & \textbf{High ripple} & \textbf{Low ripple} \\
\midrule
8 & 8.1.0.35.25 & 8.1.0.60.45 \\
\bottomrule
\end{tabular}

\end{table}

The configuration, $8.1\_0.60\_45$, exhibits a consistently low effective ripple across the plasma radius, with
$\epsilon_{\mathrm{eff}}^{3/2}$ remaining below $10^{-2}$ over a large fraction
of the core region. When compared with reference devices, this configuration exhibits ripple levels comparable to those of W7-X and LHD over a substantial radial range, while not exceeding their overall performance. These results demonstrate that an appropriate choice of TF-coil geometry can significantly reduce neoclassical transport despite the simplicity of the coil configuration.

\subsection{Configuration comparison}
\begin{figure*}[h!]
  \centering
  \includegraphics[width=0.9\textwidth,clip]{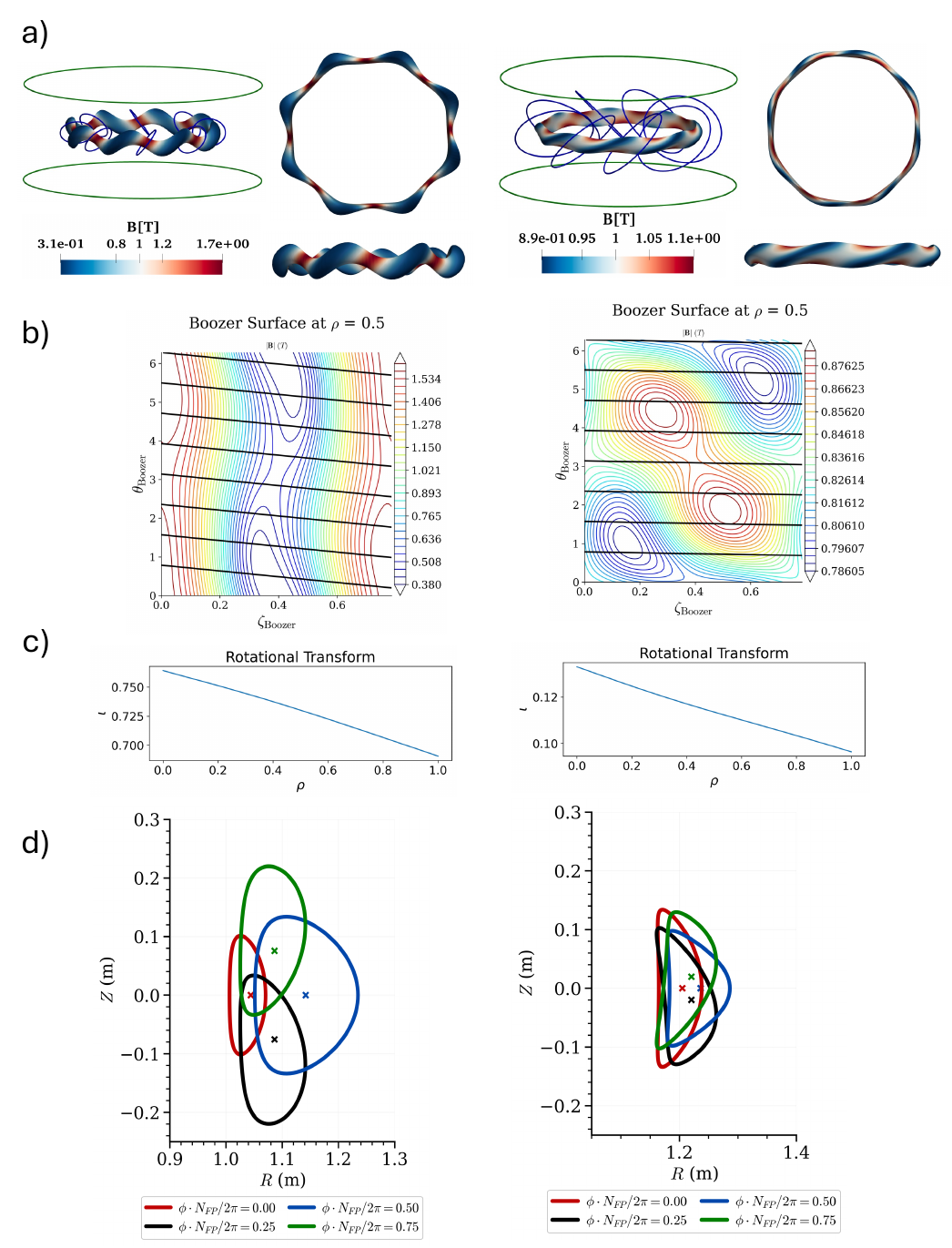}
  \caption{Comparison of the 8-TF coil configurations with high and low effective ripple, 8.1\_0.25\_35 (left) and 8.1\_0.60\_45 (right). 
(a) Three-dimensional magnetic field strength distribution on the last closed flux surface. 
(b) Contours of $|B|$ on the Boozer surface at $\rho = 0.5$. 
(c) Rotational-transform profiles as a function of normalized radius $\rho$. 
(d) Cross-sectional views of the last closed flux surfaces (LCFS) at several toroidal locations.}

  \label{fig:8TF_comp}
\end{figure*}

In this section, stellarator configurations exhibiting the lowest and 
highest effective ripple as identified from Fig.~\ref{fig:eff} listed in Table~\ref{tab:config_comparison} is compared. Figure~\ref{fig:8TF_comp} presents a comparison of the magnetic field structure and equilibrium properties of the two representative configurations with the highest and lowest effective ripple, namely 8.1\_0.25\_35 and 8.1\_0.60\_45, respectively.

\begin{table}[h!]
\centering
\medskip
\textbf{8.1\_0.6\_45}

\begin{tabular}{cccc}
\toprule
$\rho$ & $B_{\min}$ (T) & $B_{\max}$ (T) & $\Delta$ \\
\midrule
0.10 & 0.9616 & 0.9896 & 0.0144 \\
0.50 & 0.9258 & 1.0296 & 0.0531 \\
1.00 & 0.8889 & 1.1173 & 0.1139 \\
\bottomrule
\end{tabular}

\vspace{0.5cm}

\textbf{8.1\_0.25\_35}

\begin{tabular}{cccc}
\toprule
$\rho$ & $B_{\min}$ (T) & $B_{\max}$ (T) & $\Delta$ \\
\midrule
0.10 & 0.4210 & 1.5896 & 0.5812 \\
0.50 & 0.3565 & 1.6504 & 0.6447 \\
1.00 & 0.2232 & 1.9177 & 0.7915 \\
\bottomrule
\end{tabular}
\caption{: Magnetic field extrema ($B_{\max}$ and $B_{\min}$) and mirror ratio 
$\Delta = (B_{\max}-B_{\min})/(B_{\max}+B_{\min})$
for the two 8-TF configurations.}
\label{tab:mirror_ratio}
\end{table}

Figure~\ref{fig:8TF_comp}(a) shows three-dimensional visualizations of the magnetic field strength on the last closed flux surface. The high–effective-ripple configuration, $8.1\_0.25\_35$, exhibits a pronounced variation between the maximum and minimum magnetic field strength, leading to the formation of deep magnetic wells. This behavior is characteristic of linked-mirror configurations and indicates the presence of significant poloidal ripple. These observations are further quantified in Table~\ref{tab:mirror_ratio}, which lists the magnetic field extrema ($B_{\max}$ and $B_{\min}$) and the corresponding mirror ratio, $\Delta = (B_{\max} - B_{\min})/(B_{\max} + B_{\min})$, at selected normalized radii $\rho$. The configuration $8.1\_0.25\_35$ exhibits large mirror ratios even near the magnetic axis, with $\Delta = 0.5812$ at $\rho = 0.1$, increasing to $\Delta = 0.7915$ at $\rho = 1.0$. Such high values indicate strong magnetic field non-uniformity and deep trapping wells across the plasma cross-section, which directly explains the significantly elevated effective ripple and the associated increase in neoclassical transport for this configuration.In contrast, the low–effective-ripple configuration $8.1\_0.60\_45$ exhibits a much more uniform magnetic field distribution along the field lines, with a significantly smaller difference between $B_{\max}$ and $B_{\min}$, resulting in shallow magnetic wells. This behavior is consistent with the reduced effective ripple observed for this configuration. The mirror ratio remains low throughout the plasma volume, increasing gradually from $\Delta = 0.0144$ at $\rho = 0.1$ to $\Delta = 0.1139$ at the plasma edge. This moderate radial increase indicates weak magnetic field non-uniformity and minimal particle trapping, consistent with the improved neoclassical transport properties of this configuration.

Figure~\ref{fig:8TF_comp}(b) shows the magnetic field strength in Boozer coordinates at $\rho = 0.5$ for the two configurations. The high effective ripple configuration $8.1\_0.25\_35$ exhibits a strongly poloidally symmetric $|B|$ structure in Boozer coordinates, characteristic of a mirror-like magnetic configuration. In contrast, the low effective ripple configuration $8.1\_0.60\_45$ exhibits no clear symmetry in its $|B|$ structure. While this configuration has reduced effective ripple, the absence of symmetry may explain why its neoclassical transport in the $1/\nu$ regime is not better than that of fully optimized stellarators such as W7X.

Fig.~\ref{fig:8TF_comp}(c) compares the rotational transform profiles for the two configurations. The high effective ripple configuration $8.1\_0.25\_35$ exhibits a relatively large rotational transform across the plasma radius, which arises primarily from the strong poloidal ripple introduced by the smaller coil radius and lower tilt angle. While a higher rotational transform can be beneficial for stability, in this case it is achieved at the expense of increased magnetic field non-uniformity and high mirror ratio. On the other hand, the 8.1\_0.60\_45 configuration features a substantially lower rotational transform, reflecting low mirror ratio and smoother magnetic field.

Finally, Fig.~\ref{fig:8TF_comp}(d) shows cross-sectional views of the last closed flux surfaces (LCFS) at several toroidal locations. For the 8.1\_0.25\_35 configuration, the LCFS are widely spread and exhibit significant radial variation, consistent with the high rotational transform and strong magnetic field variation. In contrast, the LCFS of the 8.1\_0.60\_45 configuration remains compact and closely spaced, indicating well-defined and smoothly nested flux surfaces and reduced magnetic shear. This compactness is consistent with the lower rotational transform and low mirror ratio for this configuration.

 \begin{figure*}[h!]
  \centering
  \includegraphics[width=0.7\textwidth,clip]{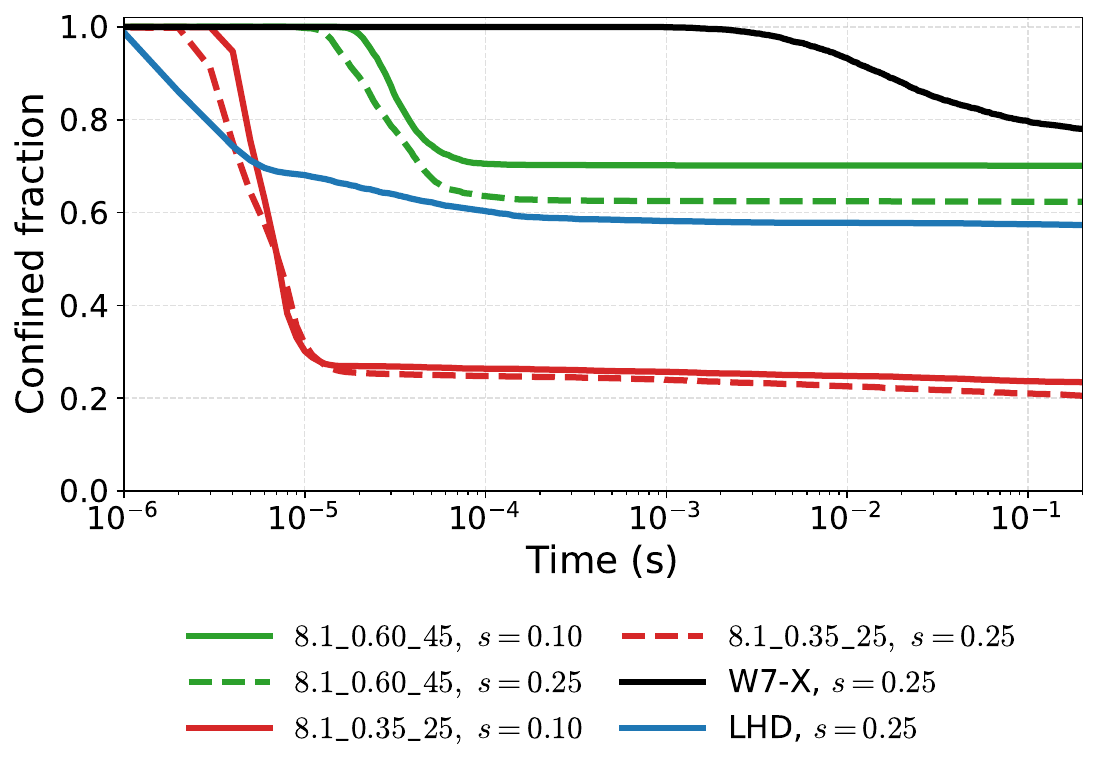}
  \caption{Time evolution of the confined fraction (y-axis) of fusion-born alpha particles for different stellarator configurations. Collisionless guiding-center trajectories are computed using the SIMPLE code by tracking 5000 alpha particles with an initial energy of $E_\alpha=3.5\,\mathrm{MeV}$ over $0.1\,\mathrm{s}$. Particles are launched from flux surfaces at $s=0.1$ and $s=0.25$ and are classified as lost upon crossing the plasma boundary at $s=1$.}
  \label{fig:alpha_particle}
\end{figure*}

\subsection{Fast particle confinement}
The confinement of fusion-born alpha particles was assessed by evaluating
collisionless losses in an ARIES-CS scale stellarator reactor configuration,
following the reactor parameters described by Najmabadi \textit{et al.}~\cite{Najmabadi08}. Collisionless guiding center trajectories were computed within the plasma volume using the symplectic gyro center code SIMPLE~\cite{Albert20}. In these simulations, radial electric fields, slowing-down processes, and wave particle interactions were neglected.
 
A total of 5000 alpha particles were initialized in the SIMPLE simulations, each with an initial energy of $E_\alpha = 3.5\,\mathrm{MeV}$, corresponding to the birth energy of alpha particles produced in D--T fusion reactions. The particles were initialized with a uniform distribution in pitch angle and are launched from magnetic flux surfaces at two radial locations, $s = 0.1$ and $s = 0.25$, which are treated independently in separate simulations. Particles are classified as lost when they cross the plasma boundary at $s = 1$. The cumulative fraction of lost fast particles was evaluated after $100\,\mathrm{ms}$ of evolution. This time interval is representative of the characteristic alpha--electron momentum exchange time at reactor-relevant plasma temperatures and densities in stellarator configurations.

Fig.~\ref{fig:alpha_particle} compares the time evolution of the confined fraction of fusion-born alpha particles for different configurations and initial launch radii. The configuration with low effective ripple (8.1\_0.60\_45) exhibits substantially improved confinement compared to the configuration (8.1\_0.25\_35) with high effective ripple at both radial locations. For particles launched near the magnetic axis ($s = 0.1$), the 8.1\_0.60\_45 configuration retains more than 65\% of the alpha particles over the full simulation duration, whereas the 8.1\_0.25\_35 configuration suffers rapid early losses, with the confined fraction dropping below 30\% within $10^{-5}\,\mathrm{s}$. A similar trend is observed for particles launched at $s = 0.25$, where enhanced radial drift leads to higher overall losses; nevertheless, the configuration 8.1\_0.60\_45 still maintains a significantly higher confined fraction than the configuration 8.1\_0.25\_35. 

\begin{figure*}[h!]
  \centering
  \includegraphics[width=0.7\textwidth,clip]{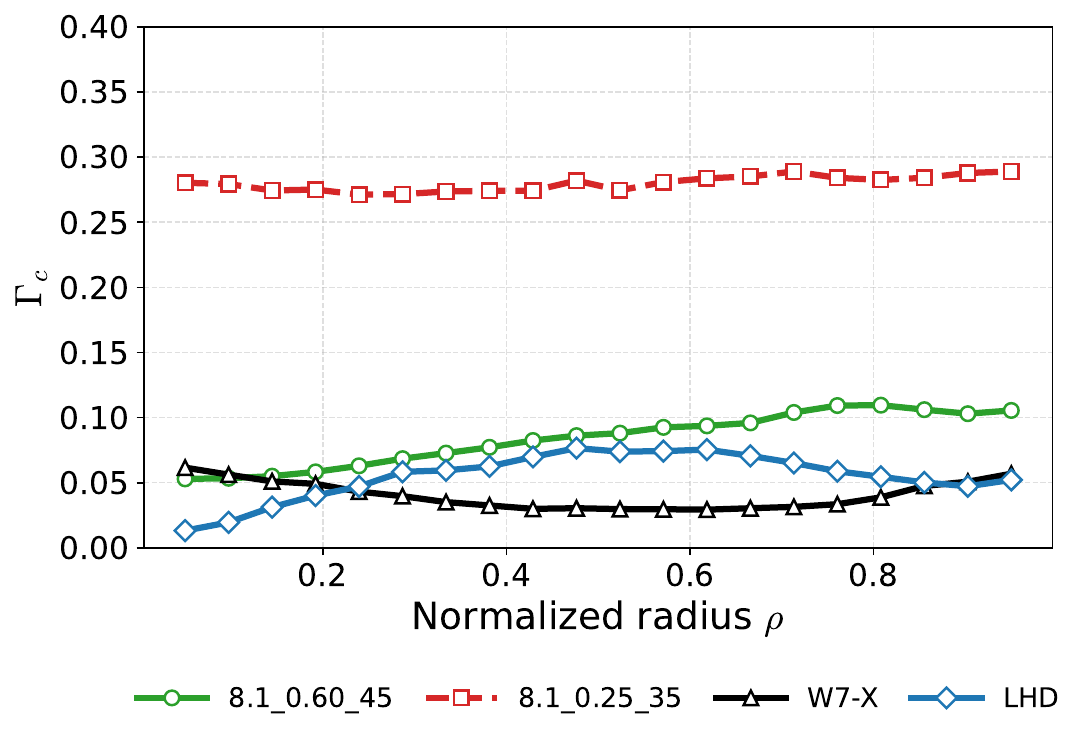}
  \caption{Radial profiles of the collisionless proxy $\Gamma_c$ as a function of normalized radius $\rho$ for the \text{8.1\_0.60\_45}, \text{8.1\_0.25\_35}, W7-X and LHD configurations.}
  \label{fig:GammaC}
\end{figure*}

The reference W7-X configuration at $s = 0.25$ shows the highest level of confinement, reflecting its fully optimized magnetic field produced by a complex modular coil set, which is inherently superior to configurations employing simpler coil geometries. Despite this, the 8.1\_0.60\_45 configuration achieves approximately 60\% alpha-particle confinement at $s = 0.25$, representing one of the highest confinement levels reported to date for stellarator configurations based on simple coil designs.

\subsection{Collisionless proxy GammaC}

In stellarators, collisionless fast-ion confinement is governed by the topology of contours of the second adiabatic invariant. Good confinement requires that contours of constant $J$ close poloidally on a given magnetic flux surface. If these contours fail to close, trapped particles experience secular radial drifts, which can lead to collisionless losses.

The well-known $\Gamma_c$ proxy~\cite{Nemov08,Bader21} is defined as
\begin{equation}
\Gamma_c(s)
=
\frac{\pi}{4\sqrt{2}}
\left\langle
\int_{B_{\max}^{-1}}^{B^{-1}}
\mathrm{d}\lambda\,
\frac{B}{\sqrt{1-\lambda B}}
\left(\gamma_c^{*}\right)^2
\right\rangle ,
\label{eq:gammac}
\end{equation}
with
\begin{equation}
\gamma_c^{*}(s)
=
\frac{2}{\pi}
\arctan
\left(
\frac{\partial_\alpha J}{\partial_s J}
\right)
=
\frac{2}{\pi}
\frac{\overline{\mathbf{v}_M \cdot \nabla s}}
{\left| \overline{\mathbf{v}_M \cdot \nabla \alpha} \right|}.
\label{eq:gammac_star}
\end{equation}
Here, $s$ is the VMEC radial coordinate (normalized toroidal flux), 
$\alpha$ is the field-line label defined as $\alpha = \theta - \iota \zeta$, 
with $\theta$ and $\zeta$ the generalized poloidal and toroidal magnetic 
coordinates, respectively; $\lambda = v_\perp^2/(B v^2)$ is the pitch-angle 
coordinate; $J$ is the second adiabatic invariant; $\mathbf{v}_M$ is the magnetic 
drift velocity; the overline denotes a bounce average, and 
$\langle \cdot \rangle$ denotes a flux-surface average.

In this formulation, $\Gamma_c$ directly measures the variation of the second adiabatic invariant with respect to the field-line label. If $\partial J / \partial \alpha = 0$, the contours of constant $J$ close poloidally and trapped particles remain confined to a flux surface. Finite values of $\partial J / \partial \alpha$ indicate broken closure of $J$ contours, leading to enhanced radial transport and collisionless fast-ion losses. Thus, small values of $\Gamma_c$ correspond to improved collisionless confinement, whereas larger values indicate degraded fast-ion confinement.

\begin{figure*}[h!]
  \centering
  \includegraphics[width=0.9\textwidth,clip]{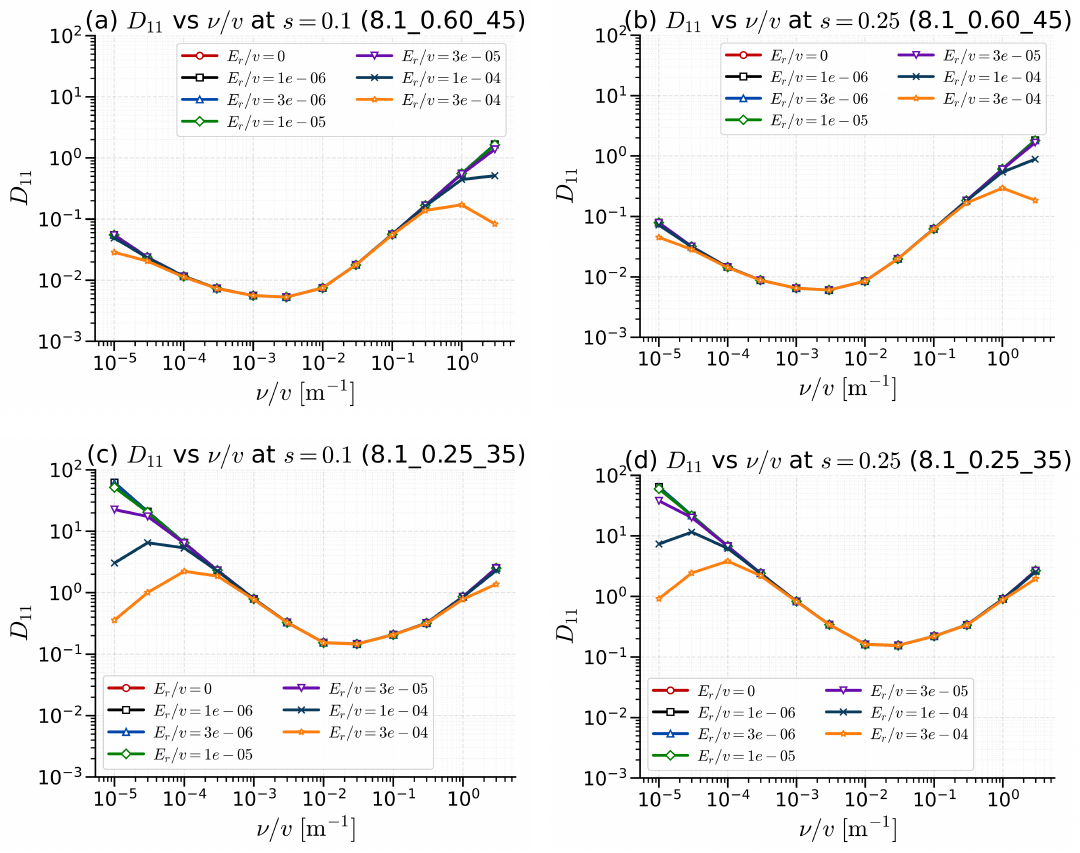}
  \caption{Mono-energetic neoclassical radial transport coefficient $D_{11}$ as a function of normalized collisionality $\nu/v$ for two stellarator configurations. The \text{8.1\_0.60\_45} configuration is shown in the top row: (a) $s=0.1$ and (b) $s=0.25$, while the \text{8.1\_0.25\_35} configuration is shown in the bottom row: (c) $s=0.1$ and (d) $s=0.25$. Results are presented for different normalized radial electric field strengths $E_r/v$, as indicated in the legends. The optimized 8\_0.60\_45 configuration exhibits substantially reduced $D_{11}$ across the entire collisionality range, particularly in the low-collisionality regime, compared to the 8\_0.25\_35 configuration, for which enhanced $1/\nu$ transport is observed.}
   \label{fig:D11}
\end{figure*}
Figure~\ref{fig:GammaC} shows the radial profile of $\Gamma_c$ for the 
configurations 8.1\_0.60\_45 and 8.1\_0.25\_35 together with the W7X and LHD. The configuration 8.1\_0.60\_45 exhibits significantly lower $\Gamma_c$ across the plasma radius, indicating 
reduced bounce-averaged radial drift and improved alignment of constant-$J$ contours with the magnetic flux surfaces. In contrast, the higher and nearly flat $\Gamma_c$ profile of the 8.1\_0.25\_35 configuration suggests a larger population of trapped particles with unfavorable drift characteristics and degraded collisionless confinement. The reduction in $\Gamma_c$ is consistent with the lower effective ripple $\epsilon_{\mathrm{eff}}$ for the 8.1\_0.60\_45 configuration, indicating improved neoclassical and fast-ion confinement, as further supported by SIMPLE simulation results.

\begin{figure*}[!h]
  \centering
  \includegraphics[width=0.58\textwidth]{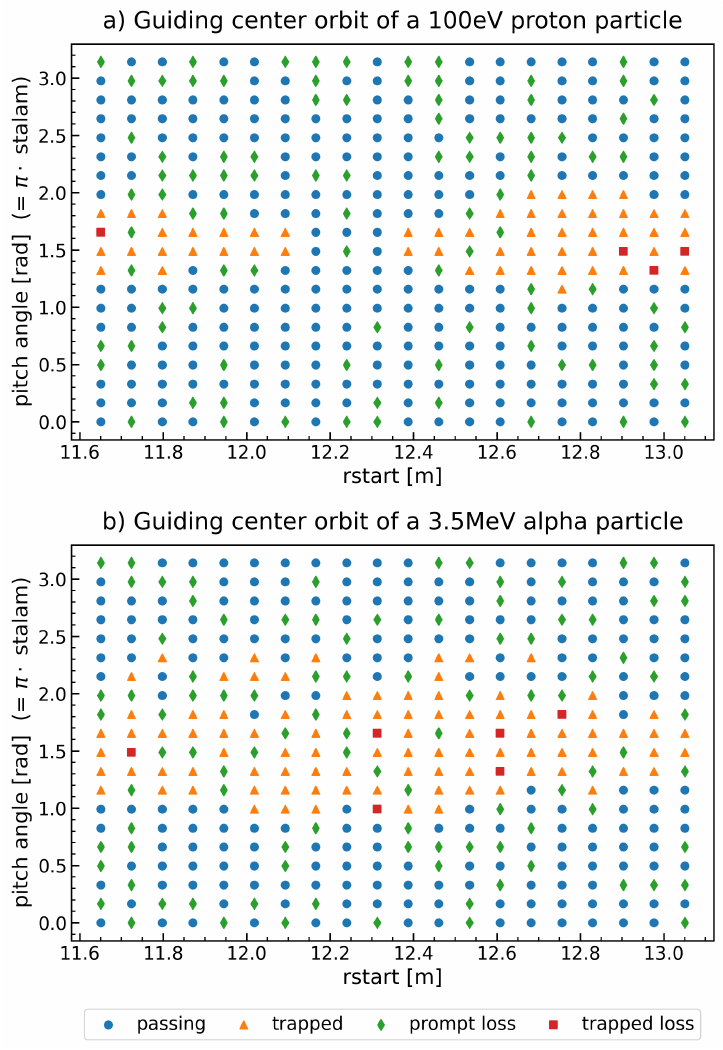}
  \caption{Guiding-center orbit classification in $(r_{\mathrm{start}},\lambda)$ space for (a) 100\,eV protons and (b) 3.5\,MeV alpha particles. Each marker represents one orbit integration and is labeled as passing (blue), trapped (orange), prompt loss (green), or trapped loss (red). The alpha particle exhibits a broader trapped and loss region than the proton, reflecting its larger orbit width and greater sensitivity to magnetic-field variations.
}
  \label{fig:gcr}
\end{figure*}

\subsection{Neoclassical transport coefficient $D_{11}$}
The mono energetic neoclassical transport coefficient $D_{11}$ quantifies radial particle transport across magnetic flux surfaces and serves as a fundamental measure of neoclassical confinement in stellarators. It represents the averaged response of guiding center radial drifts to thermodynamic gradients and is obtained as an energy-resolved moment of the solution to the linearized drift kinetic equation. As shown by Beidler \textit{et al.}~\cite{Beidler2011}, $D_{11}$ is strongly influenced by magnetic field non-axisymmetry and particle trapping, exhibiting the unfavorable $1/\nu$ scaling in the long mean free path regime due to ripple trapped particle orbits. The magnitude of $D_{11}$ is therefore closely linked to the effective ripple and serves as a sensitive indicator of magnetic optimization, with quasi-symmetric and quasi-isodynamic configurations exhibiting a substantial reduction in radial transport. The mono-energetic radial transport coefficient $D_{11}$ for the two configurations calculated using the DKES code is shown in figure~\ref{fig:D11} as a function of normalized collisionality $\nu/v$ at the radial locations $s=0.1$ and $s=0.25$. For the \text{8.1\_0.25\_35} configuration, $D_{11}$ exhibits a pronounced increase toward low collisionality, characteristic of strong $1/\nu$ transport arising from ripple-trapped particle orbits and consistent with its relatively large effective ripple $\epsilon_{\mathrm{eff}}$. In contrast, the configuration \text{8.1\_0.60\_45}  shows a substantial reduction of $D_{11}$ across the entire collisionality range at both radial locations, with particularly strong suppression in the low-collisionality regime, directly reflecting its significantly reduced $\epsilon_{\mathrm{eff}}$ and improved omnigeneity of the magnetic field. For finite normalized radial electric fields $E_r/v$, an additional reduction of $D_{11}$ is observed, especially at low collisionality where the transport departs from the $1/\nu$ regime toward electric field limited behavior; this effect is more pronounced at $s=0.25$, indicating enhanced sensitivity of outer flux surfaces. No clear plateau regime is observed, and the high-collisionality Pfirsch--Schlüter regime connects smoothly to the onset of $1/\nu$ transport at $\nu/v \sim 10^{-3}$. From a reactor perspective, the strong suppression of low-collisionality $D_{11}$ in the \text{8.1\_0.60\_45} configuration implies reduced ion and energetic-particle radial losses under reactor-relevant conditions, supporting its suitability for improved confinement in steady-state stellarator operation.

\subsection{Collisionless orbit property (single particle)}
Following the results obtained in Secs.~3.1--3.5, the configuration $8.1\_0.6\_45$ was found to be nearly comparable to optimized stellarators such as W7-X and LHD in terms of neoclassical transport and fast-ion confinement metrics. Therefore, the collisionless guiding-center orbits of a 100\,eV proton and a 3.5\, MeV alpha particle are investigated quantitatively for this configuration using the 3D Oribit Following code OFIT3D~\cite{OFIT3D}. The particles are initialized at $Z=0$ and $\phi=0$. The pitch angle, defined as $\tan^{-1}(v_\perp / v_\parallel)$, is scanned over the full range $0$ to $\pi$, and the initial radial position is varied to map the orbit topology and classify trajectories as passing, trapped, or lost. For detailed definitions of passing, trapped, trapped-loss, and prompt-loss particles, the reader is referred to Ref.~\cite{Wakatani98} (see Chapter~6). Representative passing and trapped trajectories for both the 3.5\, MeV alpha particle and the 100\,eV proton are shown in Appendix Fig.~\ref{fig:single_particle}.

Figure~\ref{fig:gcr} shows the classification of guiding-center orbits in $(r_{\mathrm{start}},\lambda)$ space for (a) 100\,eV protons and (b) 3.5\,MeV alpha particles. Each point corresponds to a single orbit integration and is classified as passing, trapped, prompt loss, or trapped loss. For the 100\,eV proton, most of the phase space is occupied by passing orbits, with trapped trajectories appearing primarily around intermediate pitch angles. Loss regions are limited and occur only in narrow bands of $(r_{\mathrm{start}},\lambda)$. In contrast, the 3.5\, MeV alpha particle exhibits a significantly broader trapped region and a larger loss domain. This behavior is consistent with its much larger Larmor radius ($r_{L,\alpha} \approx 4.6\,\mathrm{cm}$) compared to that of the proton ($r_{L,p} \approx 0.19\,\mathrm{mm}$), leading to a larger orbit width and stronger sensitivity to magnetic field variation. Consequently, alpha particles are more prone to losses than low-energy protons. Although confinement degrades with increasing particle energy, a significant fraction of phase space remains well confined for both protons and alpha particles.

\section{Conclusion}
\label{sec:conclusion}

In this study, a stellarator configuration employing tilted circular toroidal field (TF) coils was investigated, and a configuration exhibiting low effective ripple and improved alpha-particle confinement was identified through partial optimization. The optimization was performed by varying the TF coil radius and tilt angle in order to minimize the effective ripple. Two representative configurations corresponding to the highest and lowest effective ripple were compared. Although the configuration with higher effective ripple exhibits relatively good symmetry properties, it exhibits inferior alpha-particle confinement and poorer neoclassical transport performance compared to the low-effective-ripple configuration. This behavior is primarily attributed to its large mirror ratio, i.e., the significant difference between $B_{\max}$ and $B_{\min}$, which enhances magnetic trapping and increases collisionless particle losses. In contrast, the configuration with a lower effective ripple has a smaller mirror ratio, indicating reduced helical ripple and a shallower magnetic well depth, thereby improving confinement characteristics. The neoclassical transport coefficient $D_{11}$ was evaluated and found to be low for the optimized configuration. In addition, collisionless guiding-center orbit calculations for 100~eV protons and 3.5~MeV alpha particles demonstrate favorable confinement properties. However, the configuration does not exhibit strong quasi-symmetry, which limits its neoclassical transport performance and alpha-particle confinement when compared to fully optimized stellarator configurations such as W7-X. While fully optimized stellarators can achieve superior confinement and transport properties, they typically require highly complex, non-planar coil geometries. Such complexity contradicts the primary objective of the present study, which is to explore stellarator designs based on simplified coil shapes while maintaining acceptable confinement performance. The results, therefore, demonstrate a meaningful trade-off between magnetic optimization and engineering simplicity, highlighting the potential of tilted circular coil configurations as a viable compromise between physics performance and coil design complexity. The authors believe that simple-coil stellarators may, in principle, achieve alpha-particle confinement comparable to or even exceeding that of fully optimized configurations. However, the accessible parameter subspace for such simplified coil geometries is likely to be narrow. Identifying these favorable configurations requires systematic exploration, and developing robust optimization strategies tailored to simple coil sets remains an open problem. Establishing such methodologies constitutes an important direction for future research.

\clearpage
\onecolumn
\section*{Appendix}

\noindent\begin{minipage}{\textwidth}
\centering
\includegraphics[width=0.65\textwidth,clip]{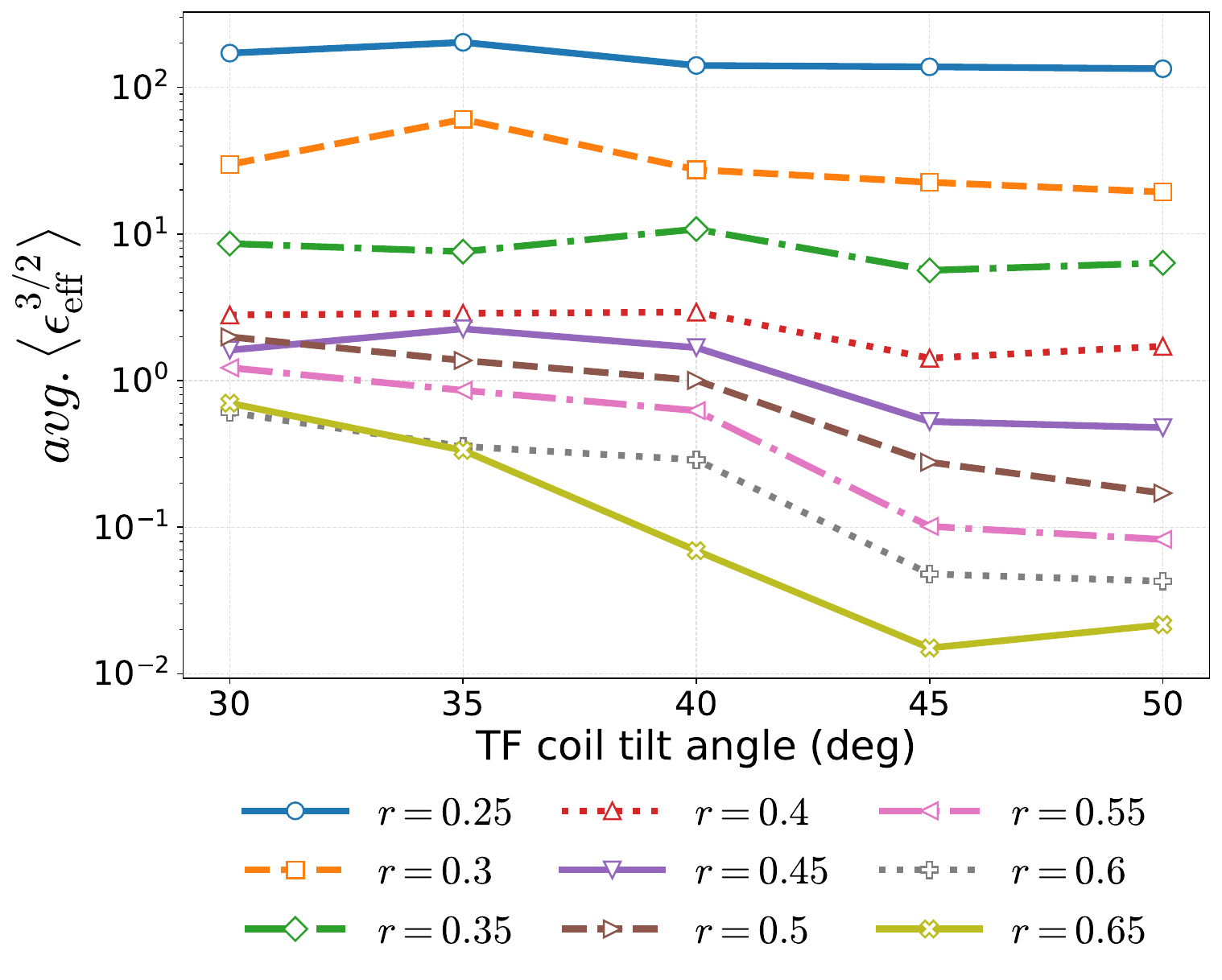}
\captionof{figure}{Variation of the flux-surface-averaged neoclassical effective ripple
$\langle \epsilon_{\mathrm{eff}}^{3/2} \rangle$ with TF coil tilt angle for the
6-NFP stellarator configuration under vacuum magnetic field conditions
($\beta = 0$). Each curve corresponds to a distinct TF coil radius $r$.}
\label{fig:6_eff}
\end{minipage}

\vspace{1em}

\noindent\begin{minipage}{\textwidth}
\centering
\includegraphics[width=0.65\textwidth,clip]{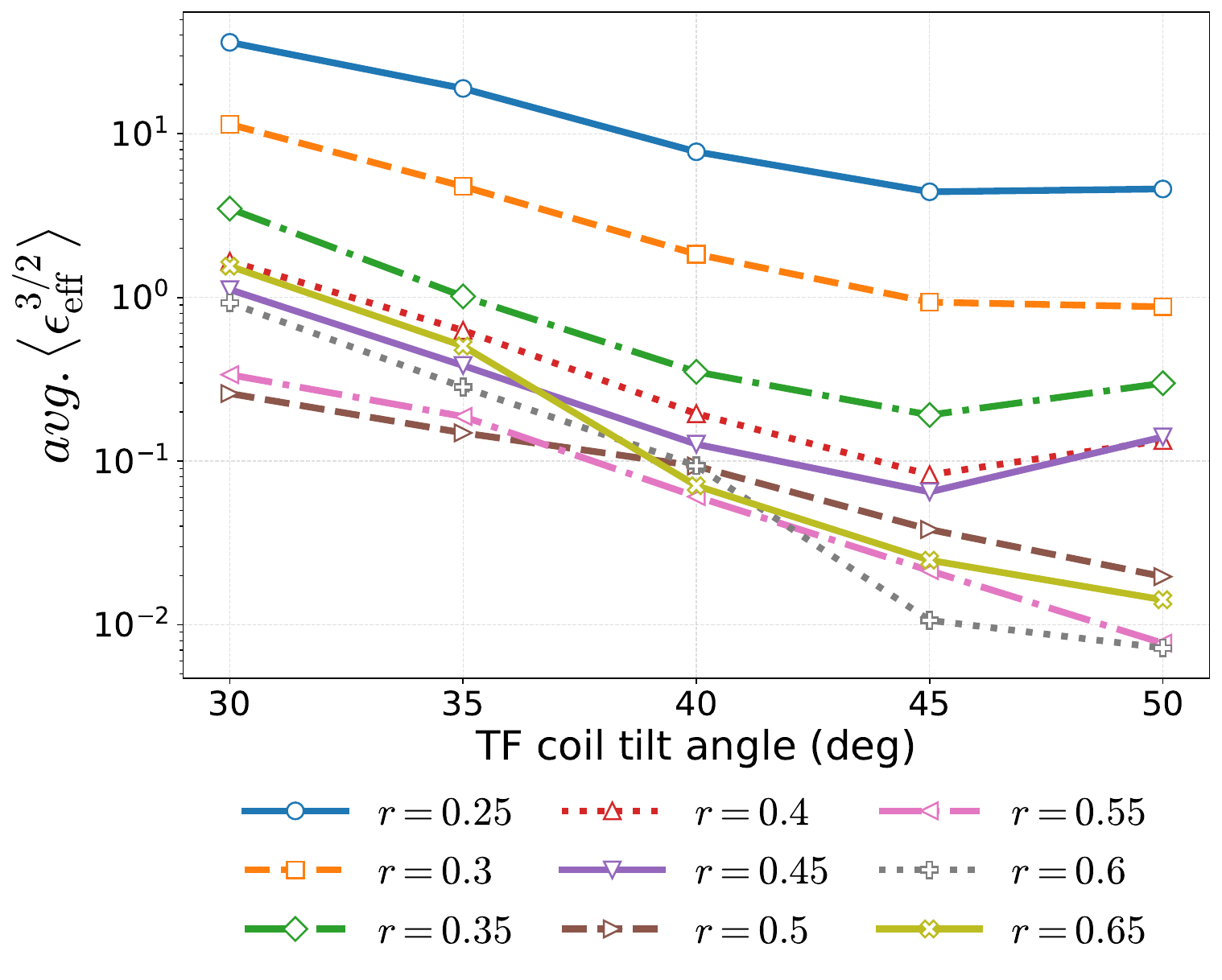}
\captionof{figure}{Variation of the flux-surface-averaged neoclassical effective ripple
$\langle \epsilon_{\mathrm{eff}}^{3/2} \rangle$ with TF coil tilt angle for the
10-NFP stellarator configuration under vacuum magnetic field conditions
($\beta = 0$). Each curve corresponds to a distinct TF coil radius $r$.}
\label{fig:10_eff}
\end{minipage}

\clearpage

\begin{figure*}[!h]
  \centering
  \includegraphics[width=0.95\textwidth]{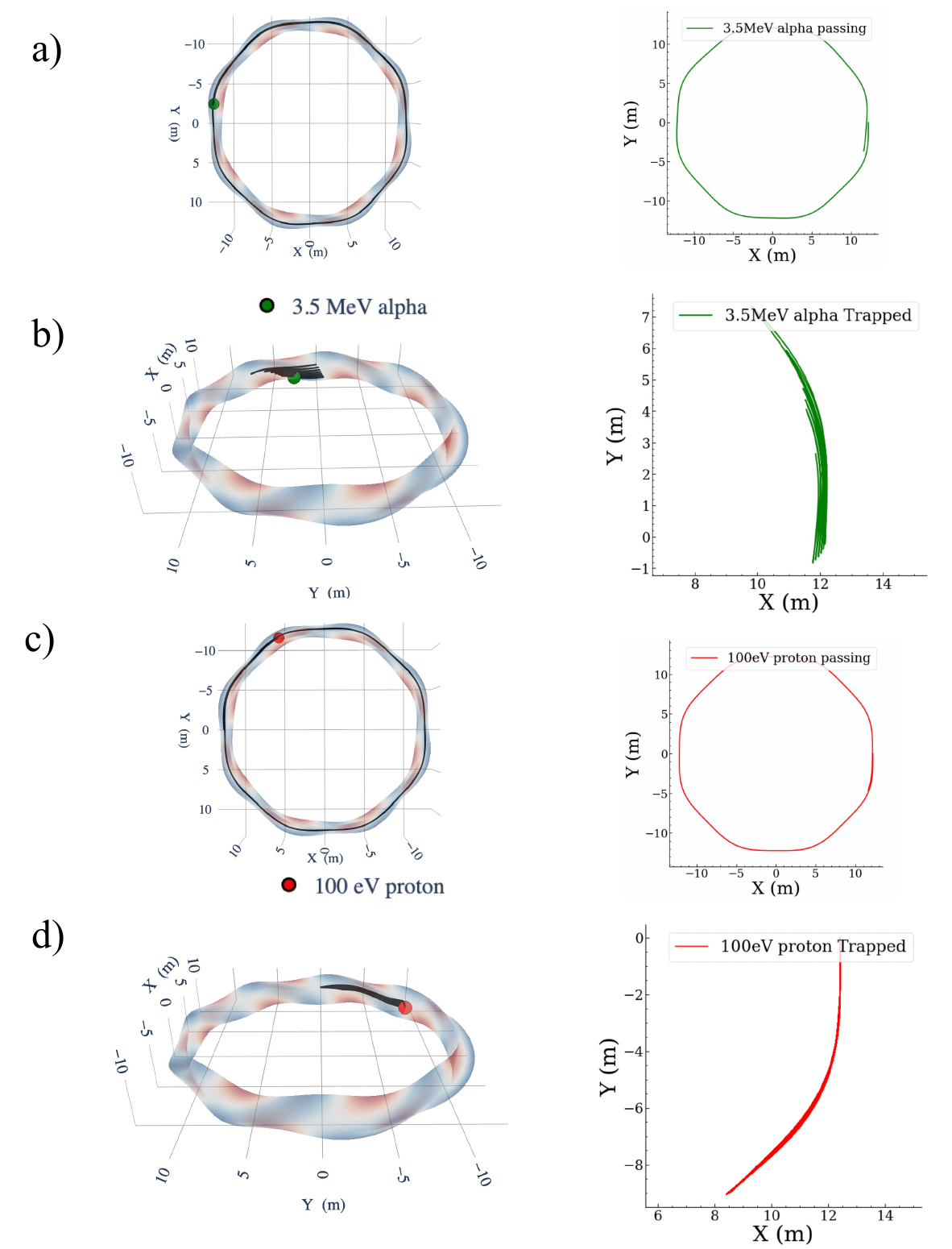}
  \caption{Single-particle guiding-center trajectories in the configuration $8.1\_0.60\_45$. (a) 3.5\,MeV alpha, passing; (b) 3.5\,MeV alpha, trapped; (c) 100\,eV proton, passing; (d) 100\,eV proton, trapped. Left panels show trajectories overlaid on the 3D plasma boundary, while right panels show $(X,Y)$ projections. Passing particles circulate toroidally, whereas trapped particles exhibit banana-type orbits due to magnetic mirror effects. The proton shows reduced orbit width and radial excursion compared to the alpha particle due to its lower energy.}
  \label{fig:single_particle}
\end{figure*}

\end{document}